\documentclass[aps,preprint,nofootinbib]{revtex4-1}
\usepackage[english]{babel} 
\usepackage[utf8]{inputenc} 

\usepackage{graphicx}
\usepackage{pgfplots}
\usepgfplotslibrary{external} 
\usepgfplotslibrary{fillbetween}
\usepackage{tikz}
\usepackage{subfig}

\usepackage{xcolor}
\usepackage{booktabs}
\usepackage{textcomp}
\usepackage{amssymb}

\usepackage{comment}

\usepackage{mathtools}

\usepackage{hyperref} 
\hypersetup{
	linkcolor=blue,
	citecolor=blue,
	urlcolor=blue
}

\pgfplotsset{
	legend image with text/.style={
		legend image code/.code={%
			\node[anchor=center] at (0.3cm,0cm) {#1};
		}
	},
}

\begin{document}
\newcommand{\abs}[1]{\left|#1\right|}

	
\title{High-Reflection Coatings for Gravitational-Wave Detectors: \textit{{\large State of The Art and Future Developments}}}

\author{Alex Amato$^{1*}$, Gianpietro Cagnoli$^1$, Maurizio Canepa$^2$, Elodie Coillet$^3$, Jerome Degallaix$^1$, Vincent Dolique$^1$, Daniele Forest$^1$, Massimo Granata$^1$, Valérie Martinez$^3$, Christophe Michel$^1$, Laurent Pinard$^1$, Benoit Sassolas$^1$, Julien Teillon$^1$}

\address{\textit{$^1$Laboratoire des Matériaux Avancés, CNRS/IN2P3, F-69622 Villeurbanne, France.\\
$^2$OPTMATLAB, Università di Genova, Via Dodecaneso 33, 16146 Genova, Italy.\\
$^3$Institut Lumière et Matière, CNRS, Université de Lyon, F-69622 Villeurbanne, France.\\
$^*$Corresponding author: a.amato@lma.in2p3.fr}}


\begin{abstract}
We report on the optical, mechanical and structural characterization of the sputtered coating materials of Advanced LIGO, Advanced Virgo and KAGRA gravitational-waves detectors. We present the latest results of our research program aiming at decreasing coating thermal noise through doping, optimization of deposition parameters and post-deposition annealing. Finally, we propose sputtered Si$_3$N$_4$ as a candidate material for the mirrors of future detectors.
\end{abstract}

\maketitle

\noindent
The high-reflecting (HR) coatings of the gravitational-wave (GW) detectors Advanced LIGO\cite{0264-9381-32-7-074001}, Advanced Virgo\cite{0264-9381-32-2-024001} and KAGRA\cite{PhysRevD.88.043007} have been deposited by the Laboratoire des Matériaux Avancés (LMA) in Lyon (Fr), where they have been the object of an extensive campaign of optical and mechanical characterization. In parallel, an intense research program is currently ongoing at the LMA, aiming at the development of low-thermal-noise optical coatings. 
The materials presented in this study are deposited by ion beam sputtering (IBS), using different coaters: a commercially available Veeco SPECTOR and the custom-developed DIBS and Grand Coater (GC). Unless specified otherwise, each coater uses different sets of parameters for the ion beam sources.
Coating refractive index and thickness are measured by transmission spectrophotometry at LMA using fused silica substrates ($\varnothing$ 1", 6 mm thick) and by reflection spectroscopic ellipsometry\cite{PRATO20112877} at the OPTMATLAB using silicon substrates ($\varnothing$ 2", 1 mm thick). Results of the two techniques are in agreement within 3\% and here are presented the average values, used to calculate coating density. Structural properties are probed by Raman scattering at the Institut Lumière Matière (ILM), using fused silica substrates. Finally, coating loss angle $\phi_c$ is measured on a Gentle Nodal Suspension\cite{doi:10.1063/1.3124800,PhysRevD.93.012007} (GeNS) system at LMA, with disk-shaped resonators of fused-silica ($\varnothing$ 2" and 3" with flats, 1 mm thick) and of silicon ($\varnothing$ 3", 0.5 mm thick).
$\phi_c$ is evaluated using the resonant method\cite{nowick1972anelastic} i.e. by measuring the ring-down time of several vibrational modes of each sample. For the $i$-th mode, it writes
\[
\phi_{i,c} = \frac{1}{D_i}\left[\phi_{i,\text{tot}}-\phi_{i,s}(1-D_i)\right]\medspace,\qquad D_i = 1 - \left(\frac{f_{i,s}}{f_{i,\text{tot}}}\right)^2\frac{m_s}{m_{\text{tot}}}\medspace,
\]
where $\phi_{i,\text{tot}}$ is the loss angle of coated disk and $\phi_{i,s}$ is the loss angle of the substrate. $D_i$ is the so-called \textit{dilution factor} which can be related to $f_{i,s}$, $f_{i,\text{tot}}$, $m_s$ and $m_{\text{tot}}$\cite{PhysRevD.89.092004}, that are the frequencies and the mass of the sample before and after the coating deposition, respectively.

\section{Standard materials in gravitational-wave interferometers}
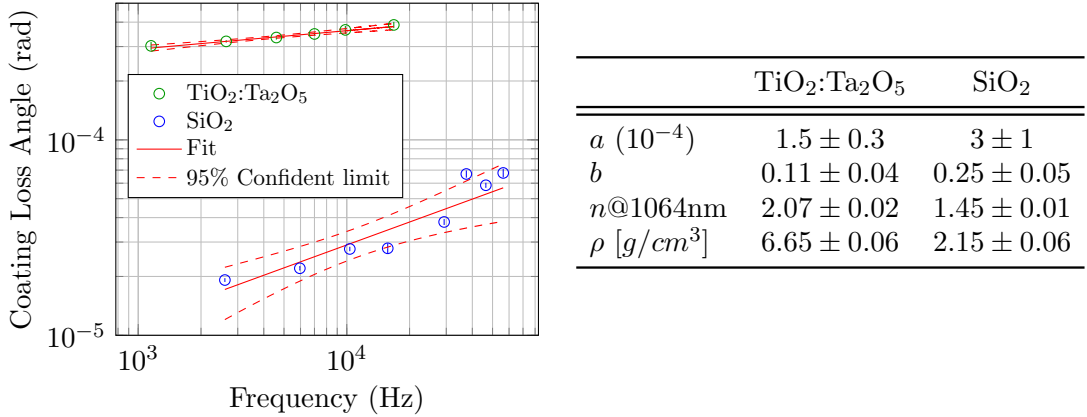
\begin{figure}[bt]  
	\centering
	\begin{tikzpicture} [x = { (1 cm , 0 cm )} , y = { (0 cm , 1 cm )}]
	\begin{axis}[
	xmode=log,
	ymode=log,
	xlabel={Frequency (Hz)},
	ylabel={Coating Loss Angle (rad)}, 
	y label style={at={(axis description cs:0,0.5)}},
	legend entries={TiO$_2$:Ta$_2$O$_5$,SiO$_2$,Fit,95\% Confident limit},
	legend style={
		cells={anchor=west},
		at={(0.68,0.6)},
		anchor=east,
		nodes={scale=0.8}
	},
	grid=both,
	ymin=0.00001,
	ymax=0.0005,
	width = 0.45\textwidth,
	height = 6cm,
	]
	\addplot [
	only marks, 
	mark = o,
	mark options={green!60!black},
	green!60!black,
	error bars/.cd, 
	error mark = none,
	y dir=both,
	y explicit
	] table [
	x = X, 
	y = Y, 
	y error = Y_error,
	] {Grafici/d14007a-c-a_freq.txt};  
	\addplot [
	only marks, 
	mark = o,
	mark options={blue},
	blue,
	error bars/.cd, 
	error mark = none,
	y dir=both,
	y explicit
	] table [
	x = X, 
	y = Y, 
	y error = Y_error,
	] {Grafici/c2f1014a-c-a_freq.txt};  
	\addplot [ 
	red,
	] table [
	x = freq, 
	y = fit,
	] {Grafici/d14007_res.txt};
	\addplot [ 
	red,
	dashed,
	] table [
	x = freq, 
	y = fit_min, 
	] {Grafici/d14007_res.txt};
	\addplot [ 
	red,
	dashed,
	] table [
	x = freq, 
	y = fit_max, 
	] {Grafici/d14007_res.txt};
	\addplot [ 
	red,
	] table [
	x = freq, 
	y = fit,
	] {Grafici/c2f1014_res.txt};
	\addplot [ 
	red,
	dashed,
	] table [
	x = freq, 
	y = fit_min, 
	] {Grafici/c2f1014_res.txt};
	\addplot [ 
	red,
	dashed,
	] table [
	x = freq, 
	y = fit_max, 
	] {Grafici/c2f1014_res.txt};
	\end{axis}
	\node at (9.5,2.3) {$\begin{array}{lcc}
		\toprule
		                & \text{TiO}_2\text{:Ta}_2\text{O}_5 & \text{SiO}_2 \\ \bottomrule\toprule
		a~(10^{-4})     &             1.5\pm0.3              &    3\pm1     \\
		b               &            0.11\pm0.04             & 0.25\pm0.05  \\
		n@1064\text{nm} &            2.07\pm0.02             & 1.45\pm0.01  \\
		\rho~[g/cm^3]   &            6.65\pm0.06             & 2.15\pm0.06  \\ \bottomrule
	\end{array}$
	};

	\end{tikzpicture}
	\caption{(Color online) Coating loss, fit parameters and optical properties of standard materials deposited by GC and annealed at 500\textcelsius~for 10 hours.}
	\label{plot:LossStand}
\end{figure}

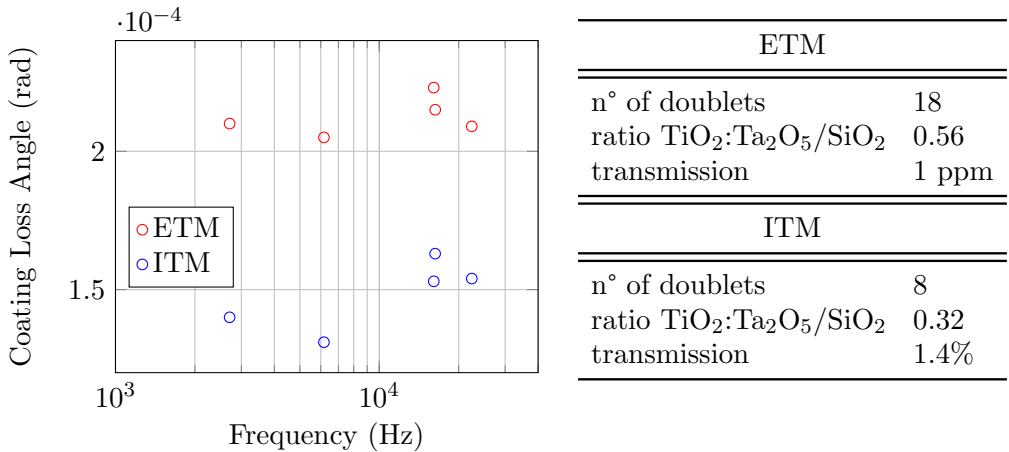
\begin{figure}[bt]  
	\centering
	\begin{tikzpicture} [x = { (1 cm , 0 cm )} , y = { (0 cm , 1 cm )}]
	\begin{semilogxaxis}[
	xlabel={Frequency (Hz)},
	ylabel={Coating Loss Angle (rad)}, 
	y label style={at={(axis description cs:0,0.5)}},
	legend entries={ETM,ITM},
	legend style={
		cells={anchor=west},
		at={(0.28,0.5)},
	},
	grid=both,
	ymin=0.00012,
	ymax=0.00024,
	xmin=1000,
	xmax=40000,
	xtick={1000,10000},
	width = 0.45\textwidth,
	height = 6cm,
	]
	\addplot [
	only marks, 
	mark = o,
	mark options={red},
	] table [
	x = Freq, 
	y = Phi, 
	] {Grafici/ETM.txt};  
	\addplot [
	only marks, 
	mark = o,
	mark options={blue},
	] table [
	x = Freq, 
	y = Phi, 
	] {Grafici/ITM.txt};  
	\end{semilogxaxis}
	\node at (9,2.3) {$\begin{array}{ll}
		\toprule
		\multicolumn{2}{c}{\text{ETM}}  \\ \bottomrule\toprule
		\text{n\textdegree~of doublets}  & 18  \\
		\text{ratio } \text{TiO}_2\text{:Ta}_2\text{O}_5/\text{SiO}_2 & 0.56 \\ 
		\text{transmission}  & 1 \text{ ppm}  \\\bottomrule\toprule
		\multicolumn{2}{c}{\text{ITM}}  \\ \bottomrule\toprule
		\text{n\textdegree~of doublets}  & 8  \\
		\text{ratio } \text{TiO}_2\text{:Ta}_2\text{O}_5/\text{SiO}_2 & 0.32 \\ 
		\text{transmission}  & 1.4\text{\%}  \\\bottomrule
	\end{array}$
	};
	\end{tikzpicture}
	\caption{(Color online) Coating loss and stack properties of ITM and ETM HR mirrors deposited by GC and annealed at 500\textcelsius~for 10 hours\cite{PhysRevD.93.012007}.}
	\label{plot:LossStack}
\end{figure}

HR coatings of Advanced LIGO and Advanced Virgo are Bragg reflectors of alternate titania-doped tantala (TiO$_2$:Ta$_2$O$_5$) and silica (SiO$_2$) layers\cite{0264-9381-32-7-074001,0264-9381-32-2-024001}. 
Fig. \ref{plot:LossStand} shows the mechanical loss of these materials, which seems to follow a power-law function $\phi_c = a\cdot f^b\medspace$.
The loss angles of the HR coatings are shown on Fig. \ref{plot:LossStack}, together with their properties. 
The end mirror (ETM) coating has higher loss angle than the input mirror (ITM) coating because of its higher ratio TiO$_2$:Ta$_2$O$_5$/SiO$_2$.
\begin{figure}[bt]  
	\centering
	\begin{tikzpicture} [x = { (1 cm , 0 cm )} , y = { (0 cm , 1 cm )}]
		\begin{axis}[
		xlabel={TiO$_2$ (\%)},
		ylabel={Coating Loss Angle (rad)}, 
		y label style={at={(axis description cs:0.08,0.5)}},
		legend style={
			cells={anchor=west},
			legend pos=north west,
			nodes={scale=0.8}
		},
		grid=both,
		minor tick num=1,
		legend entries={a),TiO$_2$:Ta$_2$O$_5$},
		width = 0.45\textwidth,
		height = 6cm,
		xmin=-5,
		xmax=105
		]
		\addlegendimage{legend image with text=}
		\addplot [ 
		only marks, 
		mark = o,
		mark options={blue},
		blue,
		error bars/.cd, 
		error mark = none,
		y dir=both,
		y explicit
		] table [
		x = X, 
		y = Y, 
		y error = eY
		] {Grafici/doping.txt};
		\end{axis}
		\end{tikzpicture}\quad
	\begin{tikzpicture} [x = { (1 cm , 0 cm )} , y = { (0 cm , 1 cm )}]
		\begin{semilogxaxis}[
		xlabel={Frequency (Hz)},
		ylabel={Coating Loss Angle (rad)}, 
		y label style={at={(axis description cs:0.08,0.5)}},
		legend style={
			cells={anchor=west},
			legend pos=north west,
			nodes={scale=0.8}
		},
		grid=both,
		legend entries={b),Ta$_2$O$_5$, TiO$_2$:Ta$_2$O$_5$},
		width = 0.45\textwidth,
		height = 6cm,
		ymin=2e-4,
		ymax=6e-4
		]
		\addlegendimage{legend image with text=}
		\addplot [ 
		only marks, 
		mark = o,
		mark options={red},
		red,
		error bars/.cd, 
		error mark = none,
		y dir=both,
		y explicit
		] table [
		x = X, 
		y = Y, 
		y error = Y_error
		] {Grafici/d14008a-c-a_freq.txt};  
		\addplot [ 
		only marks, 
		mark = o,
		mark options={green!60!black},
		green!60!black,
		error bars/.cd, 
		error mark = none,
		y dir=both,
		y explicit
		] table [
		x = X, 
		y = Y, 
		y error = Y_error
		] {Grafici/d14007a-c-a_freq_total.txt};  
		\end{semilogxaxis}
		\end{tikzpicture}
	\caption{(Color online) a) Coating loss of TiO$_2$:Ta$_2$O$_5$ as function of TiO$_2$ doping. b) Comparison of Ta$_2$O$_5$ and 18\%-doped TiO$_2$:Ta$_2$O$_5$ coating loss. All GC samples annealed at 500\textcelsius~for 10 hours (100\% TiO$_2$ coating is crystallized).}
	\label{plot:Doping}
\end{figure}
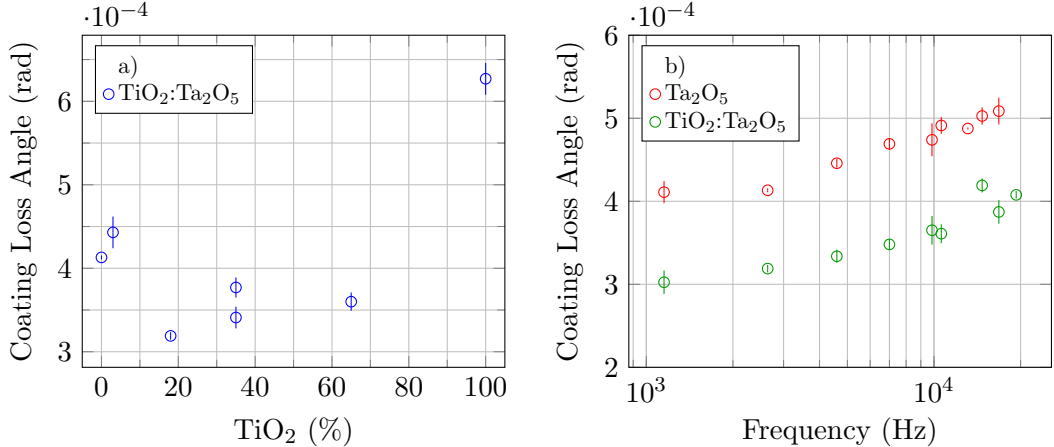 

\section{Optimization}
\subsection{Doping}
\label{sec:doping}
The purpose of TiO$_2$ doping is to increase Ta$_2$O$_5$ refractive index and reduce its loss angle. 
Increasing the refractive index contrast in the HR stack would allow to decrease the HR coating thickness, at constant reflectivity. 
Fig. \ref{plot:Doping}a shows Ta$_2$O$_5$ coating loss as function of doping. 
The current doping value in GW detectors is 18\%, which yelds a minimum loss but a refractive index only slightly higher than that of Ta$_2$O$_5$. As shown by Fig. \ref{plot:Doping}b, 18\%-doped $\phi_{\text{TiO}_2\text{:Ta}_2\text{O}_5}$ is lower than $\phi_{\text{Ta}_2\text{O}_5}$ by $\sim$25\%.
Increasing TiO$_2$ concentration will increase TiO$_2$:Ta$_2$O$_5$ refractive index, while $\phi_{\text{TiO}_2\text{:Ta}_2\text{O}_5}$ for TiO$_2\geq$40\% can not be predicted and needs further investigation.

\subsection{Deposition parameters}
\label{sec:parameters}
\begin{figure}[bt] 
	\centering
	\begin{tikzpicture} [x = { (1 cm , 0 cm )} , y = { (0 cm , 1 cm )}]
	\begin{loglogaxis}[
	xlabel={Frequency (Hz)},
	ylabel={Coating Loss Angle (rad)}, 
	y label style={at={(axis description cs:0,0.5)}},
	legend style={
		cells={anchor=west},
		legend pos=south east,
		nodes={scale=0.8}
	},
	grid=both,
	legend entries={a), Spector, GC},
	width = 0.45\textwidth,
	height = 6cm,
	ymin=1e-5,
	ymax=1e-4
	]
	\addlegendimage{legend image with text=}
	\addplot [ 
	only marks, 
	mark = o,
	mark options={green!60!black},
	green!60!black,
	error bars/.cd, 
	error mark = none,
	y dir=both,
	y explicit
	] table [
	x = X, 
	y = Y, 
	y error = Y_error
	] {Grafici/d14002a-c-a_freq_complete.txt};  
	\addplot [ 
	only marks, 
	mark = o,
	mark options={blue},
	blue,
	error bars/.cd, 
	error mark = none,
	y dir=both,
	y explicit
	] table [
	x = X, 
	y = Y, 
	y error = Y_error
	] {Grafici/c2f1014a-c-a_freq.txt};  
	\addplot [ 
	red,
	] table [
	x = freq, 
	y = fit,
	] {Grafici/c2f1014_res.txt};
	\addplot [ 
	red,
	dashed,
	] table [
	x = freq, 
	y = fit_min, 
	] {Grafici/c2f1014_res.txt};
	\addplot [ 
	red,
	dashed,
	] table [
	x = freq, 
	y = fit_max, 
	] {Grafici/c2f1014_res.txt};
	\addplot [ 
	red,
	] table [
	x = freq, 
	y = fit,
	] {Grafici/d14002_res.txt};
	\addplot [ 
	red,
	dashed,
	] table [
	x = freq, 
	y = fit_min, 
	] {Grafici/d14002_res.txt};
	\addplot [ 
	red,
	dashed,
	] table [
	x = freq, 
	y = fit_max, 
	] {Grafici/d14002_res.txt};
	\end{loglogaxis}
	\end{tikzpicture}\quad
	\begin{tikzpicture} [x = { (1 cm , 0 cm )} , y = { (0 cm , 1 cm )}]
	\begin{loglogaxis}[
	xlabel={Frequency (Hz)},
	ylabel={Coating Loss Angle (rad)}, 
	y label style={at={(axis description cs:0,0.5)}},
	legend style={
		cells={anchor=west},
		legend pos=south east,
		nodes={scale=0.8}
	},
	grid=both,
	legend entries={b), Spector (GC), GC},
	width = 0.45\textwidth,
	height = 6cm,
	ymin=1e-5,
	ymax=1e-4
	]
	\addlegendimage{legend image with text=}
	\addplot [ 
	only marks, 
	mark = o,
	mark options={green!60!black},
	green!60!black,
	error bars/.cd, 
	error mark = none,
	y dir=both,
	y explicit
	] table [
	x = X, 
	y = Y, 
	y error = Y_error
	] {Grafici/d14001a-c-a_freq.txt};  
	\addplot [ 
	only marks, 
	mark = o,
	mark options={blue},
	blue,
	error bars/.cd, 
	error mark = none,
	y dir=both,
	y explicit
	] table [
	x = X, 
	y = Y, 
	y error = Y_error
	] {Grafici/c2f1014a-c-a_freq.txt};  
	\addplot [ 
	red,
	] table [
	x = freq, 
	y = fit,
	] {Grafici/c2f1014_res.txt};
	\addplot [ 
	red,
	dashed,
	] table [
	x = freq, 
	y = fit_min, 
	] {Grafici/c2f1014_res.txt};
	\addplot [ 
	red,
	dashed,
	] table [
	x = freq, 
	y = fit_max, 
	] {Grafici/c2f1014_res.txt};
	\addplot [ 
	red,
	] table [
	x = freq, 
	y = fit,
	] {Grafici/d14001_res.txt};
	\addplot [ 
	red,
	dashed,
	] table [
	x = freq, 
	y = fit_min, 
	] {Grafici/d14001_res.txt};
	\addplot [ 
	red,
	dashed,
	] table [
	x = freq, 
	y = fit_max, 
	] {Grafici/d14001_res.txt};
	\end{loglogaxis}
	\end{tikzpicture}
	\caption{(Color online) Coating loss of SiO$_2$ deposited in the GC and in the Spector: a) different deposition parameters. b) same GC deposition parameters. All samples annealed at 500\textcelsius~for 10 hours. Fit model is the same as in Fig. \ref{plot:LossStand}.}
	\label{plot:SiO2}
\end{figure}
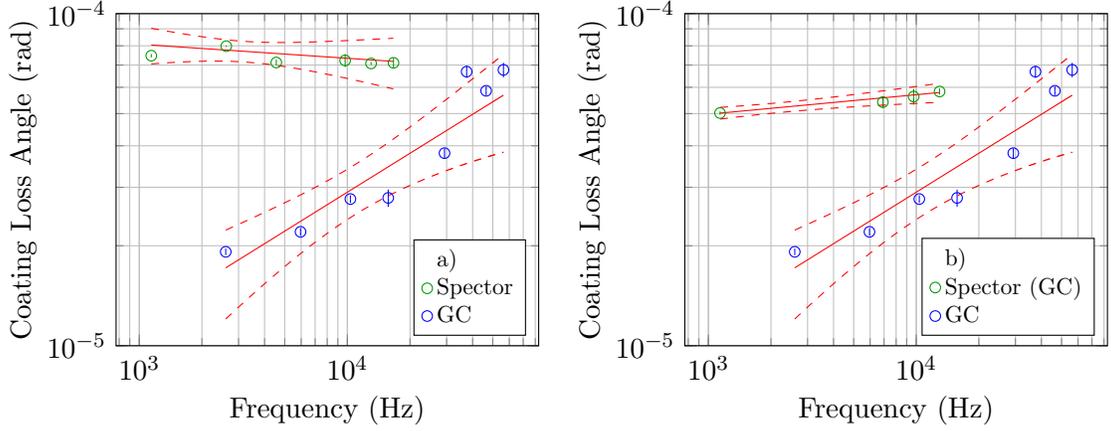

Fig. \ref{plot:SiO2}a shows coating loss of SiO$_2$ deposited by GC and Spector using their respective standard deposition parameters.
It is clear that by using different parameters the same material gets different properties: GC parameters yield lower coating loss. For further test, SiO$_2$ has been deposited in the Spector with GC parameters. As Fig. \ref{plot:SiO2}b shows, Spector coating loss is lower but still higher than GC coating loss, because of the different configuration of the coaters.
Coating losses of Ta$_2$O$_5$ deposited using different coaters have different values before annealing 
but converge toward a common limit value after, as shown in Fig. \ref{plot:LossAngleTaO}, suggesting that annealing 'deletes' the deposition history of the sample. 
Material properties are listed in table \ref{tab:fitresults}.
\begin{table}[bt]
	
	\centering
	\begin{tabular}{lcc|ccc}
		\cmidrule[1pt]{2-6} &    \multicolumn{2}{c}{SiO$_2$}        &         \multicolumn{3}{c}{Ta$_2$O$_5$}         \\
		                    &    Spector    & Spector as GC     &     GC      &   Spector   &     DIBS      \\ \bottomrule\toprule
		$a~(10^{-4})$       &   $1.1\pm0.2$   &  $0.33\pm0.01$      & $1.89\pm0.09$ &  $2.6\pm0.5$  &  $2.08\pm0.08$  \\
		$b~(10^{-1})$       &  $-0.4\pm0.3$   &  $0.59\pm0.05$      & $1.00\pm0.05$  & $0.67\pm0.08$ &  $0.88\pm0.04$  \\
		$n$            & $1.474\pm0.005$ & $1.468\pm0.005$     & $2.03\pm0.02$ & $2.08\pm0.02$ & $2.014\pm0.005$ \\
		$\rho~[g/cm^3]$     &  $2.34\pm0.01$  &  $2.33\pm0.05$      & $7.34\pm0.07$ &  $7.5\pm0.1$  &   $6.9\pm0.2$   \\ \bottomrule
	\end{tabular}
	\caption{Coating loss, fit parameters ($\phi_c = a\cdot f^b\medspace$), refractive index $n$ at $\lambda=1064$ nm and density $\rho$ of materials deposited in different coaters.}
	\label{tab:fitresults}
\end{table}
\begin{figure}[bt]  
	\centering
	\begin{tikzpicture} [x = { (1 cm , 0 cm )} , y = { (0 cm , 1 cm )}]
		\begin{axis}[
		xmode=log,
		ymode=log,
		xlabel={Frequency (Hz)},
		ylabel={Coating Loss Angle (rad)}, 
		y label style={at={(axis description cs:0,0.5)}},
		legend entries={a),0h,5h,10h,30h,70h,189h,300h},
		legend style={
			cells={anchor=west},
			at={(0.31,0.6)},
			anchor=east,
			nodes={scale=0.8}
		},
		grid=both,
		width = 0.45\textwidth,
		height = 6cm,
		ymin=0.00002,
		ymax=0.001,
		]
		\addlegendimage{legend image with text=}
		\addplot [
		only marks, 
		mark = o,
		mark options={green!60!black},
		green!60!black,
		error bars/.cd, 
		error mark = none,
		y dir=both,
		y explicit
		] table [
		x = X_0, 
		y = Y_0, 
		y error = eY_0,
		] {Grafici/Loss_SiO2_time_d14002.txt};  
		\addplot [
		only marks, 
		mark = o,
		mark options={orange},
		orange,
		error bars/.cd, 
		error mark = none,
		y dir=both,
		y explicit
		] table [
		x = X_5, 
		y = Y_5, 
		y error = eY_5,
		] {Grafici/Loss_SiO2_time_d14002.txt};  
		\addplot [
		only marks, 
		mark = o,
		mark options={blue},
		blue,
		error bars/.cd, 
		error mark = none,
		y dir=both,
		y explicit
		] table [
		x = X_10, 
		y = Y_10, 
		y error = eY_10,
		] {Grafici/Loss_SiO2_time_d14002.txt};  
		\addplot [
		only marks, 
		mark = o,
		mark options={red},
		red,
		error bars/.cd, 
		error mark = none,
		y dir=both,
		y explicit
		] table [
		x = X_30, 
		y = Y_30, 
		y error = eY_30,
		] {Grafici/Loss_SiO2_time_d14002.txt};  
		\addplot [
		only marks, 
		mark = o,
		mark options={green},
		green,
		error bars/.cd, 
		error mark = none,
		y dir=both,
		y explicit
		] table [
		x = X_70, 
		y = Y_70, 
		y error = eY_70,
		] {Grafici/Loss_SiO2_time_d14002.txt};  
		\addplot [
		only marks, 
		mark = o,
		mark options={violet},
		violet,
		error bars/.cd, 
		error mark = none,
		y dir=both,
		y explicit
		] table [
		x = X_189, 
		y = Y_189, 
		y error = eY_189,
		] {Grafici/Loss_SiO2_time_d14002.txt};  
		\addplot [
		only marks, 
		mark = o,
		mark options={purple},
		purple,
		error bars/.cd, 
		error mark = none,
		y dir=both,
		y explicit
		] table [
		x = X_305, 
		y = Y_305, 
		y error = eY_305,
		] {Grafici/Loss_SiO2_time_d14002.txt};  
		\end{axis}
		\end{tikzpicture}\quad
	\begin{tikzpicture} [x = { (1 cm , 0 cm )} , y = { (0 cm , 1 cm )}]
		\begin{axis}[
		xmode=log,
		xlabel={Frequency (Hz)},
		ylabel={Coating Loss Angle (rad)}, 
		y label style={at={(axis description cs:0,0.5)}},
		legend entries={b),0h,10h,20h},
		legend style={
			cells={anchor=west},
			at={(0.4,0.55)},
			anchor=east,
			nodes={scale=0.8}
		},
		grid=both,
		width = 0.45\textwidth,
		height = 6cm,
		ymin=0.0003,
		ymax=0.0015,
		]
		\addlegendimage{legend image with text=}
		\addplot [
		only marks, 
		mark = o,
		mark options={green!60!black},
		green!60!black,
		error bars/.cd, 
		error mark = none,
		y dir=both,
		y explicit
		] table [
		x = X_0, 
		y = Y_0, 
		y error = eY_0,
		] {Grafici/Loss_Ta2O5_time_wafer_0.txt};  
		\addplot [
		only marks, 
		mark = o,
		mark options={blue},
		blue,
		error bars/.cd, 
		error mark = none,
		y dir=both,
		y explicit
		] table [
		x = X_10, 
		y = Y_10, 
		y error = eY_10,
		] {Grafici/Loss_Ta2O5_time_wafer.txt};  
		\addplot [
		only marks, 
		mark = o,
		mark options={red},
		red,
		] table [
		x = X_20, 
		y = Y_20, 
		] {Grafici/Loss_Ta2O5_time_wafer.txt};  
		\end{axis}
		\end{tikzpicture}\\
	\begin{tikzpicture} [x = { (1 cm , 0 cm )} , y = { (0 cm , 1 cm )}]
		\begin{axis}[
		xlabel={Raman shift $(\text{cm}^{-1})$},
		ylabel={Raman intensity (a.u.)}, 
		y label style={at={(axis description cs:0.03,0.5)}},
		legend style={
			cells={anchor=west},
			nodes={scale=0.8},
			legend pos=north east,
		},
		legend entries={c),0h,5h,10h,30h,70h,184h,300h},
		legend pos = north east,
		width = 0.45\textwidth,
		height = 6cm,
		xmax=900,
		xmin=200,
		minor tick num=1,
		ytick={0,400,800,1200,1600,2000},
		yticklabels={0,0.2,0.4,0.6,0.8,1},
		ymin=0,
		ymax=2000
		]
		\addlegendimage{legend image with text=}
		\addplot [ 
		mark = none,
		thick,
		cyan,
		line width=0.5pt,
		] table [
		x = X_0, 
		y = Y_0, 
		] {Grafici/1b-S15022-NR.txt};  
		\addplot [ 
		mark = none,
		thick,
		blue,
		line width=0.5pt,
		] table [
		x = X_5, 
		y = Y_5, 
		] {Grafici/1b-S15022-5h.txt};  
		\addplot [
		mark = none,
		thick,
		green!50!black,
		line width=0.5pt,
		] table [
		x = X_10, 
		y = Y_10, 
		] {Grafici/1b-S15022-10h.txt};  
		\addplot [ 
		mark = none,
		thick,
		orange,
		line width=0.5pt,
		] table [
		x = X_30, 
		y = Y_30, 
		] {Grafici/1b-S15022-30h.txt};  
		\addplot [ 
		mark = none,
		thick,
		red,
		line width=0.5pt,
		] table [
		x = X_70, 
		y = Y_70, 
		] {Grafici/1b-S15022-70h.txt};  
		\addplot [ 
		mark = none,
		thick,
		green,
		line width=0.5pt,
		] table [
		x = X_184, 
		y = Y_184, 
		] {Grafici/1b-S15022-184h.txt};  
		\addplot [ 
		mark = none,
		thick,
		violet,
		line width=0.5pt,
		] table [
		x = X_300, 
		y = Y_300, 
		] {Grafici/1b-S15022-300h.txt};  
		\end{axis}
		\end{tikzpicture}\quad
	\begin{tikzpicture} [x = { (1 cm , 0 cm )} , y = { (0 cm , 1 cm )}]
		\begin{axis}[
		xlabel={Raman shift $(\text{cm}^{-1})$},
		ylabel={Raman intensity (a.u.)}, 
		y label style={at={(axis description cs:0.03,0.5)}},
		legend style={
			cells={anchor=west},
			nodes={scale=0.8},
			legend pos=north east,
		},
		legend entries={d),0h,10h,20h,50h},
		legend pos = north east,
		width = 0.45\textwidth,
		height = 6cm,
		xmax=1200,
		xmin=0,
		minor tick num=1,
		ytick={0,1000,2000,3000,4000,5000},
		yticklabels={0,0.2,0.4,0.6,0.8,1},
		ymin=0,
		ymax=5000
		]
		\addlegendimage{legend image with text=}
		\addplot [ 
		mark = none,
		thick,
		cyan,
		line width=0.5pt,
		] table [
		x = X_0, 
		y = Y_0, 
		] {Grafici/Raman_Ta2O5_time.txt};  
		\addplot [ 
		mark = none,
		thick,
		blue,
		line width=0.5pt,
		] table [
		x = X_10, 
		y = Y_10, 
		] {Grafici/Raman_Ta2O5_time.txt};  
		\addplot [
		mark = none,
		thick,
		green!50!black,
		line width=0.5pt,
		] table [
		x = X_20, 
		y = Y_20, 
		] {Grafici/Raman_Ta2O5_time.txt};  
		\addplot [ 
		mark = none,
		thick,
		red,
		line width=0.5pt,
		] table [
		x = X_50, 
		y = Y_50, 
		] {Grafici/Raman_Ta2O5_time.txt};  
		\end{axis}
		\end{tikzpicture}
	\caption{(Color online) Effect of annealing duration $\Delta t$\cite{granata2017correlated}. top row: a) loss of SiO$_2$, b) loss of Ta$_2$O$_5$; bottom row: c) structure of SiO$_2$, d) structure of Ta$_2$O$_5$.}
	\label{plot:AnnLosstime}
\end{figure}
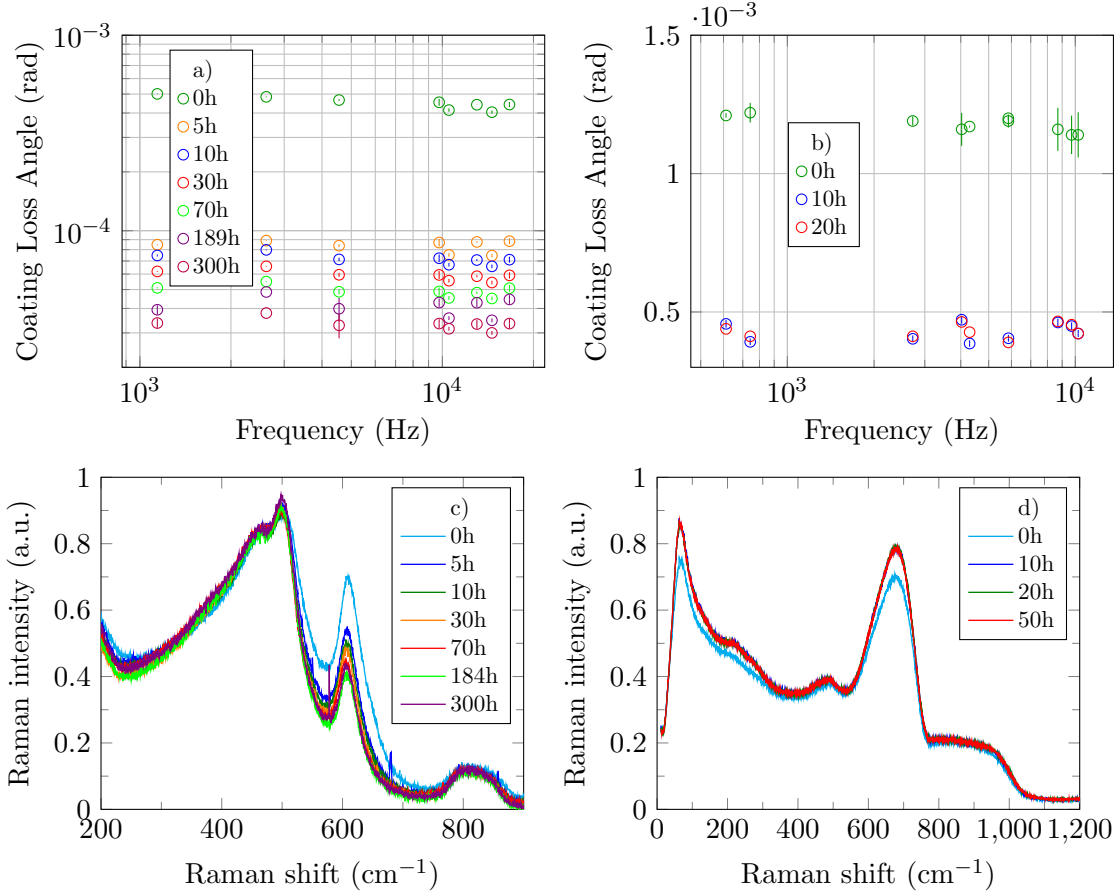
\begin{figure}[bt]  
	\centering
	\begin{tikzpicture} [x = { (1 cm , 0 cm )} , y = { (0 cm , 1 cm )}]
		\begin{axis}[
		xmode=log,
		ymode=log,
		xlabel={Frequency (Hz)},
		ylabel={Coating Loss Angle (rad)}, 
		y label style={at={(axis description cs:0,0.5)}},
		legend entries={a),0\textcelsius,500\textcelsius,900\textcelsius},
		legend style={
			cells={anchor=west},
			at={(0.31,0.75)},
			anchor=east,
			nodes={scale=0.75}
		},
		grid=both,
		width = 0.45\textwidth,
		height = 6cm,
		ymin=0.000008,
		ymax=0.0006
		]
		\addlegendimage{legend image with text=}
		\addplot [
		only marks, 
		mark = o,
		mark options={green!60!black},
		green!60!black,
		error bars/.cd, 
		error mark = none,
		y dir=both,
		y explicit
		] table [
		x = X_0, 
		y = Y_0, 
		y error = eY_0,
		] {Grafici/Loss_SiO2_temp_d14002.txt};  
		\addplot [
		only marks, 
		mark = o,
		mark options={blue},
		blue,
		error bars/.cd, 
		error mark = none,
		y dir=both,
		y explicit
		] table [
		x = X_500, 
		y = Y_500, 
		y error = eY_500,
		] {Grafici/Loss_SiO2_temp_d14002.txt};  
		\addplot [
		only marks, 
		mark = o,
		mark options={red},
		red,
		error bars/.cd, 
		error mark = none,
		y dir=both,
		y explicit
		] table [
		x = X_900, 
		y = Y_900, 
		y error = eY_900,
		] {Grafici/Loss_SiO2_temp_d14002_900.txt};  
		\end{axis}
		\end{tikzpicture}\quad
	\begin{tikzpicture} [x = { (1 cm , 0 cm )} , y = { (0 cm , 1 cm )}]
		\begin{axis}[
		xmode=log,
		xlabel={Frequency (Hz)},
		ylabel={Coating Loss Angle (rad)}, 
		y label style={at={(axis description cs:0,0.5)}},
		legend entries={b),0\textcelsius,400\textcelsius,500\textcelsius,600\textcelsius},
		legend style={
			cells={anchor=west},
			at={(0.41,0.50)},
			anchor=east,
			nodes={scale=0.8}
		},
		grid=both,
		width = 0.45\textwidth,
		height = 6cm,
		ymax=0.0015,
		ymin=0.0003
		]
		\addlegendimage{legend image with text=}
		\addplot [
		only marks, 
		mark = o,
		mark options={green!60!black},
		green!60!black,
		error bars/.cd, 
		error mark = none,
		y dir=both,
		y explicit
		] table [
		x = X_0, 
		y = Y_0, 
		y error = eY_0,
		] {Grafici/Loss_Ta2O5_temp_wafer.txt};  
		\addplot [
		only marks, 
		mark = o,
		mark options={blue},
		blue,
		error bars/.cd, 
		error mark = none,
		y dir=both,
		y explicit
		] table [
		x = X_400, 
		y = Y_400, 
		y error = eY_400,
		] {Grafici/Loss_Ta2O5_temp_wafer_400.txt};  
		\addplot [
		only marks, 
		mark = o,
		mark options={orange},
		orange,
		error bars/.cd, 
		error mark = none,
		y dir=both,
		y explicit
		] table [
		x = X_500, 
		y = Y_500, 
		y error = eY_500,
		] {Grafici/Loss_Ta2O5_temp_wafer_500.txt};  
		\addplot [
		only marks, 
		mark = o,
		mark options={violet},
		violet,
		error bars/.cd, 
		error mark = none,
		y dir=both,
		y explicit
		] table [
		x = X_600, 
		y = Y_600, 
		y error = eY_600,
		] {Grafici/Loss_Ta2O5_temp_wafer_600.txt};  
		\end{axis}
		\end{tikzpicture}\\
	\begin{tikzpicture} [x = { (1 cm , 0 cm )} , y = { (0 cm , 1 cm )}]
		\begin{axis}[
		xlabel={Raman shift $(\text{cm}^{-1})$},
		ylabel={Raman intensity (a.u.)}, 
		y label style={at={(axis description cs:0.03,0.5)}},
		legend style={
			cells={anchor=west},
			nodes={scale=0.8},
			legend pos=north east,
		},
		legend entries={c),0\textcelsius,500\textcelsius,700\textcelsius,900\textcelsius,1000\textcelsius},
		legend pos = north east,
		width = 0.45\textwidth,
		height = 6cm,
		xmax=900,
		xmin=200,
		ymin=0,
		ymax=1200,
		minor tick num=1,
		ytick={0,220,480,720,960,1200},
		yticklabels={0,0.2,0.4,0.6,0.8,1,1.2}
		]
		\addlegendimage{legend image with text=}
		\addplot [ 
		mark = none,
		thick,
		cyan,
		] table [
		x = X_0, 
		y = Y_0, 
		] {Grafici/Raman_SiO2_temp.txt};  
		\addplot [ 
		mark = none,
		thick,
		blue,
		] table [
		x = X_500, 
		y = Y_500, 
		] {Grafici/Raman_SiO2_temp_500.txt};  
		\addplot [
		mark = none,
		thick,
		green!50!black,
		] table [
		x = X_700, 
		y = Y_700, 
		] {Grafici/Raman_SiO2_temp.txt};  
		\addplot [ 
		mark = none,
		thick,
		red,
		] table [
		x = X_1000, 
		y = Y_1000, 
		] {Grafici/Raman_SiO2_temp.txt};  
		\end{axis}
	\end{tikzpicture}\quad
	\begin{tikzpicture} [x = { (1 cm , 0 cm )} , y = { (0 cm , 1 cm )}]
	\begin{axis}[
	xlabel={Raman shift $(\text{cm}^{-1})$},
	ylabel={Raman intensity (a.u.)}, 
	y label style={at={(axis description cs:0.03,0.5)}},
	legend style={
		cells={anchor=west},
		nodes={scale=0.8},
		legend pos=north east,
	},
	legend entries={d),0\textcelsius,400\textcelsius,450\textcelsius,500\textcelsius,600\textcelsius,650\textcelsius,700\textcelsius},
	legend pos = north east,
	width = 0.45\textwidth,
	height = 6cm,
	xmax=1200,
	xmin=0,
	ymin=0,
	ymax=0.15,
	minor tick num=1,
	ytick={0,0.03,0.06,0.09,0.12,0.15},
	yticklabels={0,0.2,0.4,0.6,0.8,1}
	]
	\addlegendimage{legend image with text=}
	\addplot [ 
	mark = none,
	thick,
	cyan,
	] table [
	x = X, 
	y = Y_0, 
	] {Grafici/Raman_Ta2O5_temp.txt};  
	\addplot [ 
	mark = none,
	thick,
	blue,
	] table [
	x = X, 
	y = Y_400, 
	] {Grafici/Raman_Ta2O5_temp.txt};  
	\addplot [
	mark = none,
	thick,
	green!50!black,
	] table [
	x = X, 
	y = Y_450, 
	] {Grafici/Raman_Ta2O5_temp.txt};  
	\addplot [ 
	mark = none,
	thick,
	brown,
	] table [
	x = X, 
	y = Y_500, 
	] {Grafici/Raman_Ta2O5_temp.txt};  
	\addplot [ 
	mark = none,
	thick,
	violet,
	] table [
	x = X, 
	y = Y_600, 
	] {Grafici/Raman_Ta2O5_temp.txt};  
	\addplot [ 
	mark = none,
	thick,
	orange,
	] table [
	x = X, 
	y = Y_650, 
	] {Grafici/Raman_Ta2O5_temp.txt};  
	\addplot [ 
	mark = none,
	thick,
	red,
	] table [
	x = X, 
	y = Y_700, 
	] {Grafici/Raman_Ta2O5_temp.txt};  
	\end{axis}
	\end{tikzpicture}
	\caption{(Color online) Effect of annealing temperature $T_a$. top row: a) loss of SiO$_2$, b) loss of Ta$_2$O$_5$; bottom row: c) structure of SiO$_2$, d) structure of Ta$_2$O$_5$.}
	\label{plot:Anntemp}
\end{figure}
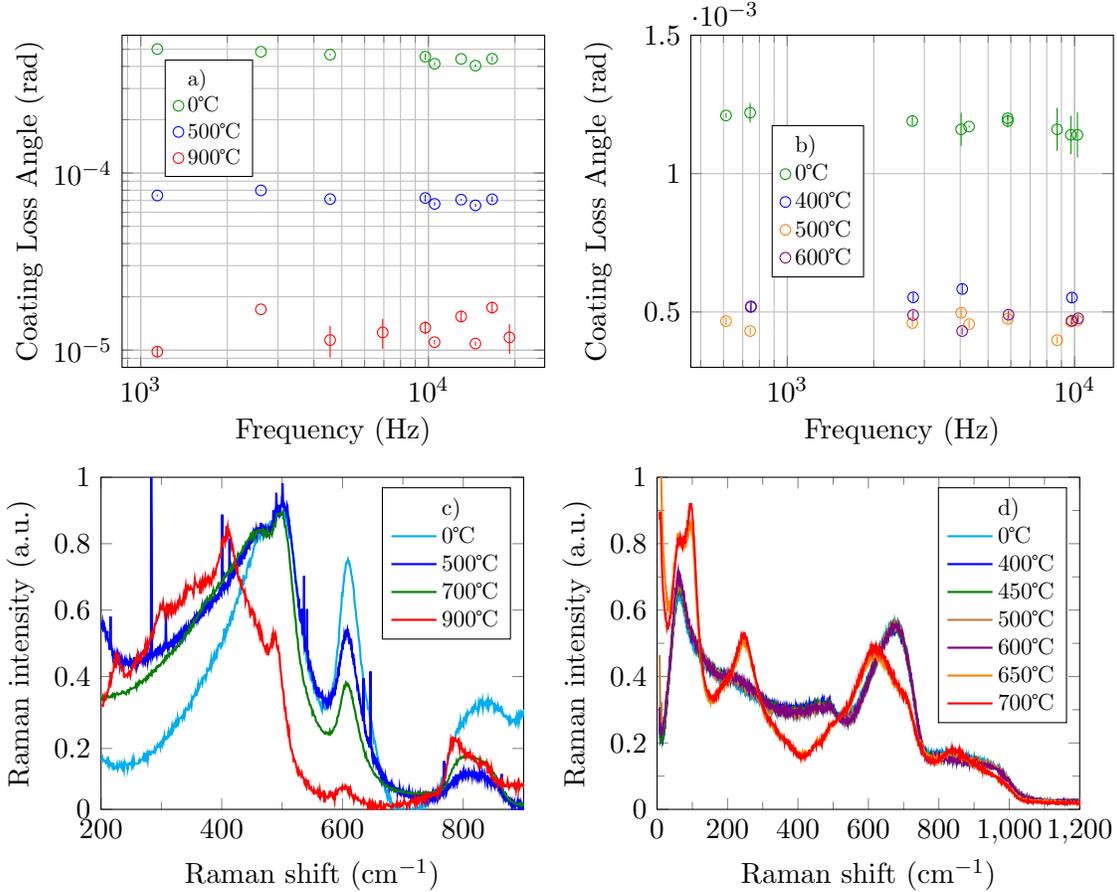 

\begin{figure}[bt]  
\begin{minipage}[c]{.45\textwidth}
\centering
\begin{tikzpicture} [x = { (1 cm , 0 cm )} , y = { (0 cm , 1 cm )}]
\begin{semilogxaxis}[
xlabel={Frequency (Hz)},
ylabel={Coating Loss Angle (rad)}, 
y label style={at={(axis description cs:0.03,0.5)}},
legend entries={Spector,GC,DIBS},
legend style={
	cells={anchor=west},
	legend pos=north west,
},
grid=both,
width = 0.85\textwidth,
height = 6cm,
]
\addplot [
only marks, 
mark = o,
mark options={green!60!black},
green!60!black,
error bars/.cd, 
error mark = none,
y dir=both,
y explicit
] table [
x = X, 
y = Y, 
y error = Y_error,
] {Grafici/c2f1019a-c-a_freq.txt};  
\addplot [
only marks, 
mark = o,
mark options={blue},
blue,
error bars/.cd, 
error mark = none,
y dir=both,
y explicit
] table [
x = X, 
y = Y, 
y error = Y_error,
] {Grafici/d14008a-c-a_freq.txt};  
\addplot [
only marks, 
mark = o,
mark options={red},
red,
error bars/.cd, 
error mark = none,
y dir=both,
y explicit
] table [
x = X, 
y = Y, 
y error = Y_error,
] {Grafici/c21006a-c-a_freq.txt};  
\end{semilogxaxis}
\end{tikzpicture}
\captionsetup{width=.9\textwidth}
\caption{(Color online) Loss of Ta$_2$O$_5$ coatings deposited by different coaters and annealed at 500\textcelsius~for 10h.}
\label{plot:LossAngleTaO}
\end{minipage}
\begin{minipage}[c]{.45\textwidth}
\centering
\begin{tikzpicture} [x = { (1 cm , 0 cm )} , y = { (0 cm , 1 cm )}]
\begin{axis}[
xlabel={Coating Loss Angle (rad)},
ylabel={Normalized D$_2$ area (a.u.)}, 
y label style={at={(axis description cs:0.05,0.5)}},
legend entries={Spector,GC,Fused Silica},
legend style={
	cells={anchor=west},
	nodes={scale=0.8},
	legend pos=south east,
},
grid=both,
width = 0.85\textwidth,
height = 6cm,
minor tick num=1,
ytick={2,4,6,8,10,12,14},
every x tick scale label/.style={
	at={(1,0)},xshift=1pt,anchor=south west,inner sep=0pt},
ymin=2,
ymax=14,
xmin=-0.000015,
xmax=0.0005
]
\addplot [
only marks, 
mark = o,
mark options={green!60!black},
green!60!black,
error bars/.cd, 
error mark = none,
y dir=both,
y explicit,
x dir=both,
x explicit
] table [
x = X,
x error = eX, 
y = Y, 
y error = eY,
] {Grafici/LossStruct.txt};  
\addplot [
only marks, 
mark = o,
mark options={blue},
blue,
error bars/.cd, 
error mark = none,
y dir=both,
y explicit,
x dir=both,
x explicit
] table [
x = X, 
x error = eX,
y = Y, 
y error = eY,
] {Grafici/LossStruct_GC.txt};  
\addplot [
only marks, 
mark = o,
mark options={red},
red,
error bars/.cd, 
error mark = none,
	y dir=both,
	y explicit,
	x dir=both,
	x explicit
	] table [
	x = X, 
	x error = eX,
	y = Y, 
	y error = eY,
	] {Grafici/LossStruct_fusedSiO.txt};  
	\end{axis}
	\end{tikzpicture}
	\captionsetup{width=.9\textwidth}
	\caption{(Color online) Correlation between D$_2$ area and loss in SiO$_2$.}
	\label{plot:Loss-struct}
\end{minipage}
\end{figure}
\subsection{Post-deposition annealing}
Annealing parameters are of fundamental importance for the purpose of reducing coating thermal noise. The problem is to find the optimal annealing temperature $T_a$ and duration $\Delta t$, avoiding coating crystallization which would increase optical loss by scattering and absorption. 
In Fig. \ref{plot:AnnLosstime} is shown the effect of increasing $\Delta t$ with $T_a=$500\textcelsius~constant. 
SiO$_2$ loss decreases and this behaviour has a structural counterpart. SiO$_2$ is composed of tetrahedral units arranged in rings of different size\cite{PhysRevB.50.118} and the area of the D$_2$ band near 600 $cm^{-1}$ is associated to 3-fold ring population\cite{PhysRevLett.80.5145}. 
A correlation between coating loss and D$_2$ has been found, suggesting that SiO$_2$ loss increases with the 3-fold ring population\cite{granata2017correlated}.
This correlation holds for different kinds of SiO$_2$, coating and bulk (Fig. \ref{plot:Loss-struct}).
On the other hand, Ta$_2$O$_5$ loss does not change for $\Delta t\geq$10h and its structure evolves only for $\Delta t\leq$10h. 
Fig. \ref{plot:Anntemp} shows coating loss and structure for increasing $T_a$, with $\Delta t=$10h constant. 
SiO$_2$ loss decreases and its structure evolves considerably. Surprisingly, crystallization occurs at $T_a=$1000\textcelsius. For Ta$_2$O$_5$, coating loss is roughly constant for $T_a>$500\textcelsius~and its structure does not change up to $T_a=$600\textcelsius, when crystallization occurs. 
\begin{figure}[bt] 
	\centering
	\begin{tikzpicture} [x = { (1 cm , 0 cm )} , y = { (0 cm , 1 cm )}]
	\begin{axis}[
	xmode = log,
	xtick = {1000,10000},
	xlabel={Frequency (Hz)},
	ylabel={Coating Loss Angle (rad)}, 
	y label style={at={(axis description cs:0.03,0.5)}},
	xmin=1000,
	legend style={
		cells={anchor=west},
		nodes={scale=0.65},
		at={(0.58,0.8)},
	},
	legend entries={0\textcelsius,500\textcelsius,700\textcelsius,800\textcelsius,900\textcelsius,TiO$_2$Ta$2$O$5$ @500\textcelsius},
	grid=both,
	width = 0.45\textwidth,
	height = 6cm,
	]
	\addplot [
	only marks, 
	mark = o,
	mark options={blue},
	blue,
	error bars/.cd, 
	error mark = none,
	y dir=both,
	y explicit
	] table [
	x = X, 
	y = Y, 
	y error = err_Y,
	] {Grafici/0_SiN_Data_mean_phiC.txt};  
	\addplot [
	only marks, 
	mark = o,
	mark options={green!60!black},
	green!60!black,
	error bars/.cd, 
	error mark = none,
	y dir=both,
	y explicit
	] table [
	x = X, 
	y = Y, 
	y error = err_Y,
	] {Grafici/0_SiN_Data_mean_phiC_a.txt};  
	\addplot [
	only marks, 
	mark = o,
	mark options={orange},
	orange,
	error bars/.cd, 
	error mark = none,
	y dir=both,
	y explicit
	] table [
	x = X, 
	y = Y, 
	y error = err_Y,
	] {Grafici/0_SiN_Data_mean_phiC_a2.txt};  
	\addplot [
	only marks, 
	mark = o,
	mark options={red!80!black},
	red!80!black,
	error bars/.cd, 
	error mark = none,
	y dir=both,
	y explicit
	] table [
	x = X, 
	y = Y, 
	y error = err_Y,
	] {Grafici/0_SiN_Data_mean_phiC_a3.txt};  
	\addplot [
	only marks, 
	mark = o,
	mark options={violet},
	violet,
	error bars/.cd, 
	error mark = none,
	y dir=both,
	y explicit
	] table [
	x = X, 
	y = Y, 
	y error = err_Y,
	] {Grafici/0_SiN_Data_mean_phiC_a4.txt};  
	\addplot [ 
	cyan,
	name path=A,
	] table [
	x = freq, 
	y = fit_min, 
	] {Grafici/d14007_res.txt};
	\addplot [ 
	cyan,
	name path=B,
	] table [
	x = freq, 
	y = fit_max, 
	] {Grafici/d14007_res.txt};
	\addplot[area legend,cyan!40] fill between[of=A and B];
	\end{axis}
	\end{tikzpicture}
	\captionsetup{width=.5\textwidth}
	\caption{(Color online) Si$_3$N$_4$ loss for different $T_a$, compared to loss of TiO$_2$:Ta$_2$O$_5$ annealed at $T_a=$500\textcelsius.}
	\label{plot:LossAngleSiN}
\end{figure}
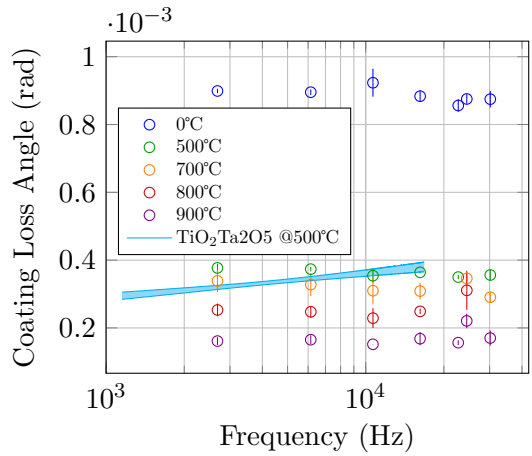
\section{New material}
TiO$_2$:Ta$_2$O$_5$ could be replaced by a material with lower mechanical loss and possibly higher refractive index.
Here silicon nitride (Si$_3$N$_4$) is proposed, which features high refractive 
index\cite{Chao2017} and very low mechanical loss\cite{Liu2007}. Usually, Si$_3$N$_4$ is deposited by low pressure chemical vapour deposition (LPCVD). However, LPCVD Si$_3$N$_4$ might suffer from hydrogen contamination and thickness uniformity issues, which are not compatible with the stringent optical specifications required for GW detectors. Instead, IBS Si$_3$N$_4$ can be developed in the GC for deposition on large optics. Fig. \ref{plot:LossAngleSiN} shows a comparison between TiO$_2$:Ta$_2$O$_5$ and IBS Si$_3$N$_4$, this latter being annealed at different temperatures.
Si$_3$N$_4$ loss decreases significantly at $T_a=$900\textcelsius. Thus, one could increase the annealing temperature of the entire HR stack to decrease also SiO$_2$ loss, eventually reducing the coating loss of the whole HR stack.

\section{Conclusions}
Coating materials of all present GW-detectors have been extensively characterized, showing a frequency-dependent loss angle. 
These standard materials can be optimized in different ways. The first approach is to increase the TiO$_2$ content in TiO$_2$:Ta$_2$O$_5$. Another option is to work on deposition parameters, in order to tune the optical and mechanical properties of the materials. In particular, the current GC configuration seems particularly well suited to deposit low loss SiO$_2$. 
In the case of Ta$_2$O$_5$, the effect of different deposition parameters is erased by the first few hours of post-deposition annealing. For 400\textcelsius$<T_a<$600\textcelsius~and 10h$<\Delta t<$50h Ta$_2$O$_5$ loss shows null or limited evolution and its structure is frozen in a stable configuration.
In the case of SiO$_2$, the annealing parameters $T_a$ and $\Delta t$ have a significant impact on mechanical loss and coating structure. A correlation is found between D$_2$ peak area in the Raman spectra, associated to the three fold ring population, and mechanical loss.
IBS Si$_3$N$_4$ is an interesting new possibility to replace TiO$_2$:Ta$_2$O$_5$ because  of its low mechanical loss. Furthermore, Si$_3$N$_4$ can be annealed at higher temperature than TiO$_2$:Ta$_2$O$_5$, reducing also SiO$_2$ coating loss angle and thus the loss of the whole HR coating.

\section*{References}
\bibliographystyle{iopart-num}
\bibliography{bibliography.bib}{}

\providecommand{\newblock}{}
\begin{thebibliography}{10}
\expandafter\ifx\csname url\endcsname\relax
  \def\url#1{{\tt #1}}\fi
\expandafter\ifx\csname urlprefix\endcsname\relax\def\urlprefix{URL }\fi
\providecommand{\eprint}[2][]{\url{#2}}

\bibitem{0264-9381-32-7-074001}
Aasi J {\em et~al.\/} 2015 {\em Classical and Quantum Gravity\/} {\bf 32}
  074001.

\bibitem{0264-9381-32-2-024001}
Acernese F {\em et~al.\/} 2015 {\em Classical and Quantum Gravity\/} {\bf 32}
  024001.

\bibitem{PhysRevD.88.043007}
Aso Y, Michimura Y, Somiya K, Ando M, Miyakawa O, Sekiguchi T, Tatsumi D and
  Yamamoto H (The KAGRA Collaboration) 2013 {\em Phys. Rev. D\/} {\bf 88}(4)
  043007

\bibitem{PRATO20112877}
Prato M, Chincarini A, Gemme G and Canepa M 2011 {\em Thin Solid Films\/} {\bf
  519} 2877 -- 2880 ISSN 0040-6090 5th International Conference on
  Spectroscopic Ellipsometry (ICSE-V)

\bibitem{doi:10.1063/1.3124800}
Cesarini E, Lorenzini M, Campagna E, Martelli F, Piergiovanni F, Vetrano F,
  Losurdo G and Cagnoli G 2009 {\em Rev. Sci. Instrum.\/} {\bf 80} 053904.

\bibitem{PhysRevD.93.012007}
Granata M {\em et~al.\/} 2016 {\em Phys. Rev. D\/} {\bf 93}(1) 012007.

\bibitem{nowick1972anelastic}
Nowick A and Berry B 1972 {\em Anelastic relaxation in crystalline solids\/}
  (Academic Press)

\bibitem{PhysRevD.89.092004}
Li T {\em et~al.\/} 2014 {\em Phys. Rev. D\/} {\bf 89}(9) 092004.

\bibitem{granata2017correlated}
Granata M {\em et~al.\/} 2017 {\em arXiv preprint arXiv:1706.02928.\/}

\bibitem{PhysRevB.50.118}
Jin W, Kalia R~K, Vashishta P and Rino J~P 1994 {\em Phys. Rev. B\/} {\bf
  50}(1) 118--131.

\bibitem{PhysRevLett.80.5145}
Pasquarello A and Car R 1998 {\em Phys. Rev. Lett.\/} {\bf 80}(23) 5145--5147.

\bibitem{Chao2017}
Chao S 2017 {\em LIGO-G1700304.\/}

\bibitem{Liu2007}
Liu X, Metcalf T~H, Wang Q and Photiadis D~M 2007 {\em MRS Proceedings\/} {\bf
  989.}

\end{thebibliography}

\end{document}